\documentclass[11pt,a4paper]{article}
\pdfoutput = 1
\usepackage{jcappub}
\usepackage[utf8]{inputenc}
\usepackage{amsmath,amssymb}
\usepackage{graphicx}
\usepackage{hyperref}
\usepackage{natbib}
\usepackage{mathtools}

\newcommand{\bp}{\textbf{p}}
\newcommand{\bq}{\textbf{q}}
\newcommand{\bx}{\textbf{x}}
\newcommand{\bk}{\textbf{k}}

\newcommand{\bs}{\textbf{s}}

\newcommand{\third}{\frac{1}{3}}

\newcommand{\avg}[1]{\ensuremath{\left\langle #1 \right\rangle}}

\renewcommand{\vec}{\mathbf}
\newcommand*{\df}  {\delta}
\newcommand*{\pp}  {\parallel}

\newcommand*{\lb}  {\left(}
\newcommand*{\rb}  {\right)}
\newcommand*{\la}  {\left\langle}
\newcommand*{\ra}  {\right\rangle}
\newcommand{\eq}[1]{\begin{align}#1\end{align}}
\newcommand{\eeq}[1]{\begin{equation}#1\end{equation}}
\DeclarePairedDelimiter\floor{\lfloor}{\rfloor}

\def\hq{\hat{q}} 
\def\hn{\hat{n}}
\def\hk{\hat{k}} 
\def\hp{\hat{p}}

\author[a]{Shi-Fan Chen}
\author[b,c]{Zvonimir Vlah}
\author[a]{Martin White}

\affiliation[a]{Department of Physics, University of California,
Berkeley, CA 94720}
\affiliation[b]{Kavli Institute for Cosmology, University of Cambridge, Cambridge CB3 0HA, UK.}
\affiliation[c]{Department of Applied Mathematics and Theoretical Physics, University of Cambridge, Cambridge CB3 0WA, UK.}

\emailAdd{shifan\_chen@berkeley.edu}
\emailAdd{zv217@cam.ac.uk}
\emailAdd{mwhite@berkeley.edu}

\title{The Ly$\alpha$ forest flux correlation function: a perturbation theory perspective}


\abstract{The Ly$\alpha$ forest provides one of the best means of mapping large-scale structure at high redshift, including our tightest constraint on the distance-redshift relation before cosmic noon.  We describe how the large-scale correlations in the Ly$\alpha$ forest can be understood as an expansion in cumulants of the optical depth field, which itself can be related to the density field by a bias expansion.  This provides a direct connection between the observable and the statistics of the matter fluctuations which can be computed in a systematic manner.  We discuss the way in which complex, small-scale physics enters the predictions, the origin of the much-discussed velocity bias and the `renormalization' of the large-scale bias coefficients.  Our calculations are within the context of perturbation theory, but we also make contact with earlier work using the peak-background split.  Using the structure of the equations of motion we demonstrate, to all orders in perturbation theory, that the large-scale flux power spectrum becomes the linear spectrum times the square of a quadratic in the cosine of the angle to the line of sight.  Unlike the case of galaxies, both the isotropic and anisotropic pieces receive contributions from small-scale physics.}

\begin{document}
\maketitle
\flushbottom


\section{Introduction}

The Ly$\alpha$ forest refers to the structure imprinted in the spectra of high redshift galaxies and quasars by absorption of photons by neutral hydrogen along the line of sight \cite{Pea99,Meiksin09,McQuinn16a}.  Most of the signal comes from warm gas near mean density that is in photoionization equilibrium with an almost uniform ultra-violet background and as such probes density fluctuations on scales ranging from the Jeans scale ($\mathcal{O}(100\,\mathrm{kpc})$) to the size of the survey (up to Gpc).  It is currently our best probe of large-scale structure from sub-Mpc to hundreds of Mpc scales at high redshift, where galaxy redshift surveys are sparse and cover limited area and intensity mapping surveys have yet to report detections.

There are two regimes in which the Ly$\alpha$ forest provides critical constraints on our cosmological models.  The first is at small scales, where the sensitivity of the forest to sub-Mpc scales allows constraints on the spectral index, massive neutrinos, warm and fuzzy dark matter and other candidates that suppress small-scale power \cite{Croft99,Viel04,Armengaud17,Baur17,Murgia18,PD20,Enzi20,Viel13,Irsic17,Irsic17b,Rogers2021}.  The second, which is the focus of this work, is on large scales where the Ly$\alpha$ forest can be thought of as a biased tracer of the density field, much like galaxies or QSOs \cite{Slosar11,Bourboux20}.  On the very largest scales\footnote{Throughout we shall neglect large-scale effects arising from general-relativistic corrections \cite{Irsic16,Lepori20} or fluctuations in the ultraviolet background field \cite{Croft04,Meiksin04,McQuinn11b,Gontcho14,Pontzen14a,Pontzen14b,Meiksin19}, focusing instead on dynamical non-linearities and the way in which small- and large-scale physics couple into the observable flux.} it has been argued that the flux, averaged over a large region, should depend linearly on the matter overdensity and peculiar velocity gradient.  In this limit the flux correlation function, or power spectrum, can be computed analytically.  The question then arises on what scales such an approximation is valid, how do non-linearities enter, to what extent do the numerous complex physical processes affecting the intergalactic medium on small scales modify the large-scale behavior and how does one systematically extend a linear theory calculation?

Traditionally models of the Ly$\alpha$ forest power spectrum addressing the above questions have involved fits to, or forms inspired by, numerical simulations (see e.g.\ refs.\ \cite{Borde14,Arinyo15,Walther21} for recent examples and the reviews cited above for a more complete overview).  In large part this is because the mapping between the underlying density fluctuations and the observed flux is highly non-linear, mixing small- and large-scale effects, which complicates an analytic or perturbative treatment (though see refs.\ \cite{Gnedin1998,McDonald00,Seljak12,Cieplak16,Irsic18,Givans20,Garny21} for some notable analytic treatments).  However the 3D power spectrum of the Ly$\alpha$ forest is observed to be very close to a linearly biased tracer of the linear power spectrum on large scales \cite{Bourboux20}, which suggests that the mixing of small- and large-scale physics does not alter the flux power spectrum beyond recognition.  In this paper we develop a systematic means of computing the flux correlation function beyond linear theory.  We show how the effects of this non-linear mapping can be encapsulated into a set of nuisance terms, and isolate the parts of the Ly$\alpha$ forest power spectrum and correlation function that come from large scales and are expected to be amenable to analytic or perturbative treatment and those which depend sensitively on small-scale physics (i.e.\ ``astrophysics'').  We discuss how these effects arise in the 1-loop flux power spectrum and demonstrate the approach to linear theory with a specific $\mu$ dependence on large scales, including the manner in which higher-order effects enter.  We demonstrate how small-scale physics impacts the bias parameters and how redshift-space distortions induce anisotropy in the clustering.  Our approach utilizes the fact that the non-linear mapping relating density to flux is of a known form (an exponential of a biased tracer of the density field), and thus should prove to be more general than the particular application we highlight here.

Earlier work \cite{Seljak12,Cieplak16} also studied the consequences of dynamical nonlinearities under exponential maps, specifically within the peak-background split.  These approaches needed to treat redshift-space distortions approximately.  Our calculations expand upon these by considering a more general set of nonlinearities within an effective-theory framework, but we make connection with these previous results where appropriate (especially in the context of real space).  We also discuss the structure of the theory beyond the 1-loop level, and the allowed angular dependence in the large-scale limit.
Recently, refs.~\cite{Givans20,Des18} proposed a second-order anisotropic bias basis in the context of Ly$\alpha$ and a third-order one for generic biased tracers with line-of-sight selection effects.  We show how the combination of dynamics and the exponential map generates the terms in those expansions.

The outline of the paper is as follows.  Section \ref{sec:formalism} shows how the transmitted flux correlation function, $\xi_F$, is related to cumulants of the optical depth fluctuation, $\delta\tau$.  Since $\delta\tau$ can be modeled in a manner quite similar to galaxy or quasar density fields in perturbation theory, this establishes the key connection between our observable and objects which are under good theoretical control.  We validate the connection between $\xi_F$ and the cumulants in Section \ref{sec:numerics} using some highly idealized simulations that illustrate our key points.  How the cumulants of $\delta\tau$ behave, how non-linear scales enter and the impact of redshift-space distortions is the topic of Section \ref{sec:structure}.  In this section we present several approaches so as to bring out the key physics and make connection with earlier work.  We conclude in Section \ref{sec:conclusions}.  Details of our calculations and explicit formulae are given in a series of appendices.

\section{Formalism}
\label{sec:formalism}

\subsection{Optical depth and density}

The relationship between the redshift-space optical depth, $\tau$, and the dark matter density, $\delta$, and velocity (divergence) $\theta$ is set by the physics of the Ly$\alpha$ forest \cite{Pea99,Meiksin09,McQuinn16a}.  Specifically the low-density gas is in photoionization equilibrium with an almost uniform ultraviolet background field, and thus traces a power of the dark matter density field on large scales, where pressure forces are unimportant.  As shown in Fig.~7 of ref.~\cite{Lukic15}, we are interested primarily in gas near the cosmic mean density.  Instead of attempting to model the complex physics determining the optical depth field, even approximately, we shall take the approach that $\tau$ can be expanded as a series in the (smoothed) density and velocity fields containing all terms allowed by the symmetries \cite{Des16}.  The coefficients of the terms in such an expansion, the bias coefficients, are to be treated as parameters of the theory that must be fit to observations or numerical simulations that resolve the small-scale physics of the forest.

Using an Eulerian biasing prescription, we have
\begin{equation}
    \tau(\bx) = \tau_0\ \left[ 1 + b_1 \delta(\bx) + \frac{1}{2} b_2 \left(\delta^2(\bx) - \avg{\delta^2}\right) + \cdots \right] + \epsilon(\bx)
\label{eqn:bias_expansion}
\end{equation}
in real space.  Here $\epsilon$ is the ``stochastic'' contribution to $\tau$ that is uncorrelated with large scales and has zero mean.  Note that in the limit of an isothermal gas where $\tau\propto (1+\delta)^\gamma$ with $\gamma=2$ this bias expansion would be exact, while for $\gamma\ne 2$ we have $b_2=\gamma(\gamma-1)$ and there would be higher order terms.  The $\cdots$ includes terms depending upon the shear field and other invariants that can be formed at higher order \cite{Des16}. Throughout this work we will work primarily in terms of fluctuations in the optical depth, defined as $\tau = \tau_0 (1 + \delta \tau)$, where $\tau_0 = \avg{\tau}$ is the mean optical depth.  Since optical depth is conserved in mapping from real to redshift space, assuming large-scale velocities are gravitationally dominated and invoking the equivalence principle $\delta\tau(\bs) = b_1 \delta + \theta$ to lowest order, with $\theta$ the velocity divergence.  We shall consider how this form is modified by higher-order corrections when going to flux in the following.

In the above discussion we have implicitly worked within the approximation where baryons and cold dark matter are treated as a single fluid. In fact, the two species can be subject to both different small-scale forces due to galactic physics (feedback, outflows, star formation etc.) \cite{Lew15} as well as subtle differences in their densities and velocities post-recombination.  These differences enter at linear order and beyond in the statistics of biased tracers \cite{Ang15,Schmidt16,Chen19} and can be especially significant at the high redshifts probed by Ly$\alpha$ measurements \cite{Tse10,Yoo11}. A systematic treatment of these terms in the context of the Ly$\alpha$ --- and similar considerations for the inclusion of massive neutrinos --- is beyond the scope of this work, though we refer readers to ref.~\cite{Givans20} who consider a subset of these operators directly proportional to the baryon-CDM relative velocity.

\subsection{Flux correlators and the exponential map}

We are interested in correlators of the continuum normalized flux field\footnote{Throughout we shall neglect observational issues such as continuum fitting, high column density systems, metal lines, etc.  The reader is referred to ref.~\cite{Bourboux20} for a discussion of these issues and references to the observational literature.}, or transmission fraction, which is defined as the ratio of the observed flux to the continuum flux when there is no intervening absorption.  We write the flux, $F$, in terms of an optical depth
\eeq{
  F(\vec x) = e^{- \tau (\vec x)}
}
and introduce fluctuations about the mean in the usual manner
\eeq{
  F(\vec x) = F_0 \lb 1 + \df F(\vec x) \rb
}
where the fluctuations are defined to have zero mean so that $\la F(\vec x) \ra = F_0$.
The mean flux, $F_0$, can be related to the cumulants of the $\tau$ field evaluated at a single point, specifically
\eeq{
  F_0 = \la F(\vec x) \ra 
  =  \la e^{- \tau (\vec x)} \ra
  =  \exp \Bigg[ \sum_{n=1}^\infty \frac{(-1)^n}{n!} \la \lb \tau (\vec x) \rb^n \ra_c \Bigg]
}
and of course $F_0$ is independent of the chosen point, $\mathbf{x}$.
Note that the relation between $F_0$ and $\tau_0$ depends upon the full PDF of $F$, and involves non-perturbative physics.  It could be measured directly, predicted from numerical simulations or left as a degree of freedom in the model.  If we write $F_0=e^{-\tau_0}(1+\delta F_0)$ then by Jensen's inequality we have $\delta F_0\ge 0$.

Next let us consider the 2-point function of the flux field, which is given by
\eeq{
  \la F(\vec x) F(\vec x') \ra = F_0^2 \left( 1 + \la \delta F(\vec x) \delta F(\vec x') \ra \right)
  = \la e^{- \tau (\vec x) - \tau (\vec x') } \ra.
}
Using the previous relation of $F_0$ to the point correlators of $\tau$ we can write (see Appendix \ref{app:details})
\eeq{
  \ln \Big( 1 + \la \delta F \delta F' \ra \Big)
  = \sum_{n=2}^\infty \tau_0^n
  \Bigg[ \frac{1+(-1)^n}{2 (n/2)!}  \xi^{(n/2,n/2)} (\vec r)
  + 2 \frac{(-1)^n}{n!} \sum_{m=1}^{\floor{\frac{n-1}{2}}} \binom{n}{m} \xi^{(m,n-m)} (\vec r) \Bigg]
\label{eqn:main}
}
which expresses the flux correlation function in terms of correlators of the optical depth fluctuations, $\xi_\tau^{(ij)}(\vec r)=\langle \delta\tau^i \delta\tau^j\rangle_c$.  These cumulants encode the manner in which the fluctuations of the optical depth on large scales (that can be modeled e.g.\ using conventional perturbation theory) affect large-scale flux correlations. Note that at a fixed value of $F_0$ the value of $\tau_0$ is sensitive to small-scale physics. As in the bias expansion, one should treat $\tau_0$ and $F_0$ as numbers that cannot be determined from large-scale physics alone -- they must be fit from data or simulations -- but the functional dependence of $\la \delta F \delta F' \ra$ is not free.  The large-scale physics (e.g.\ the baryon acoustic peak) resides in the $\xi_\tau^{(ij)}$.  We give some examples of cumulants in the following.

We note that the structure of Eq.~(\ref{eqn:main}) is dictated by the form of the relationship between $F$ and $\tau$, rather than the physics of the Ly$\alpha$ forest or cosmological dynamics.  We thus expect it to be of much more general utility than simply modeling the Ly$\alpha$ forest.  While it is that modeling that we investigate further in the rest of the paper, let us end this section by mentioning one (slight) generalization of the general formalism towards cross-correlation of the Ly$\alpha$ forest with other biased\footnote{The auto-correlation of galaxies or QSOs is already routinely modeled with perturbative methods so such measurements could be naturally included in our formalism.} tracers (e.g.\ quasars or dampled Ly$\alpha$ systems), i.e.\  $\avg{(1+\delta_Q) e^{-\tau}} = F_0\left(1+\xi_{\times}\right)$. By promoting the cumulant expansion in Eq.~(\ref{eqn:main}) to include two random variables (in this case $\delta \tau$ and $\delta_Q$) it is possible to show that
\begin{equation}
  \left\langle \delta^n e^{-\tau} \right\rangle
    = F_0  \sum_{m} \frac{(-1)^m}{m!}\ \kappa_{m,n}
  \quad , \quad
  F_0 = \exp\left[ \sum_m\frac{(-1)^m}{m!}\ \kappa_{m,0} \right]
  \quad .
\end{equation}
where $\kappa_{m,n}(\textbf{r})$ are joint cumulants of the two relative overdensities (see Appendix \ref{app:cross_correlations}).  If we use this formula with the normal bias expansions for $\delta_Q$ and $\delta\tau$ the form of $\xi_\times$ quickly follows.

\section{Numerical experiments}
\label{sec:numerics}

\subsection{Simulations}

To test the convergence of the moment and cumulant expansions we use a suite of N-body simulations.  A detailed modeling of the Ly$\alpha$ forest would require high-resolution hydrodynamic simulations, and we are not in a position to run a large volume of such simulations in order to study the large-scale behavior of the flux correlations.  Instead we have chosen a toy model which has some (though definitely not all) of the properties of the Ly$\alpha$ problem.  In particular, our focus is on the exponential mapping involved in going from $\delta\tau$ to $F$, and the behavior of the cumulant expansion.  We shall artificially increase the pressure smoothing present in the Ly$\alpha$ forest so that the moments and correlators of the mock flux field will be well behaved and we can run very large volumes with low computational cost.

To this end we employ the suite of simulations described in ref.\ \cite{Modi20} to construct mock optical depth and flux fields. These are 10 FastPM boxes with sidelength 1.536 $h^{-1}$ Gpc run with $2048^3$ particles. To produce mock optical depths we take the matter density field at $z = 2$ smoothed with a Gaussian of $R = 5$ or $10\,h^{-1}$Mpc and transformed them via the FGPA approximation \cite{Meiksin09,McQuinn16a}
\begin{equation}
    \tau(\bx) = A \left[ 1 + \delta_m(\bx) \right]^\gamma,
\end{equation}
with $\gamma = 1.5$ and $A$ chosen such that $F_0 = 0.7$.  For these simulations $\tau_0\approx 0.37 $ so that $e^{-\tau_0}\simeq 0.69$ and $\delta F_0\simeq 0.01$.  Note that our smoothing scale is roughly two orders of magnitude larger than the actual Jeans scale in the Ly$\alpha$ forest.  We want to use such a large smoothing because the correlators in our expansions become increasingly UV-sensitive and numerically ill-behaved with increased powers of $\tau$ and we wanted to ensure numerical convergence within our limited computational resources. In addition, these simulations have significantly larger volume (though lower resolution) than typical Ly$\alpha$ simulations so that we can better understand the large-scale behavior of the Ly$\alpha$ signal and series convergence.  While these simulations contain qualitatively similar nonlinearities and nonlinear maps, it was not our intent that they be faithful imitations of Ly$\alpha$ physics.

\subsection{Results}

\begin{figure}
\resizebox{0.48\textwidth}{!}{\includegraphics{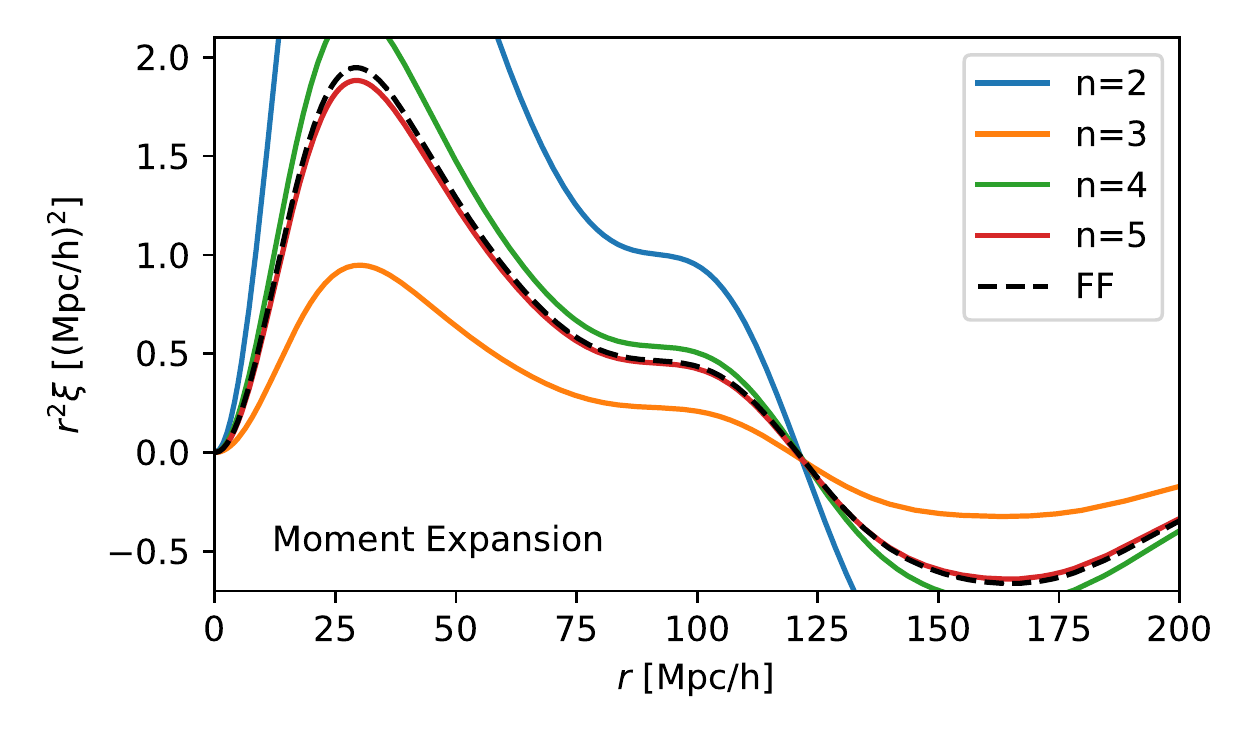}}
\resizebox{0.48\textwidth}{!}{\includegraphics{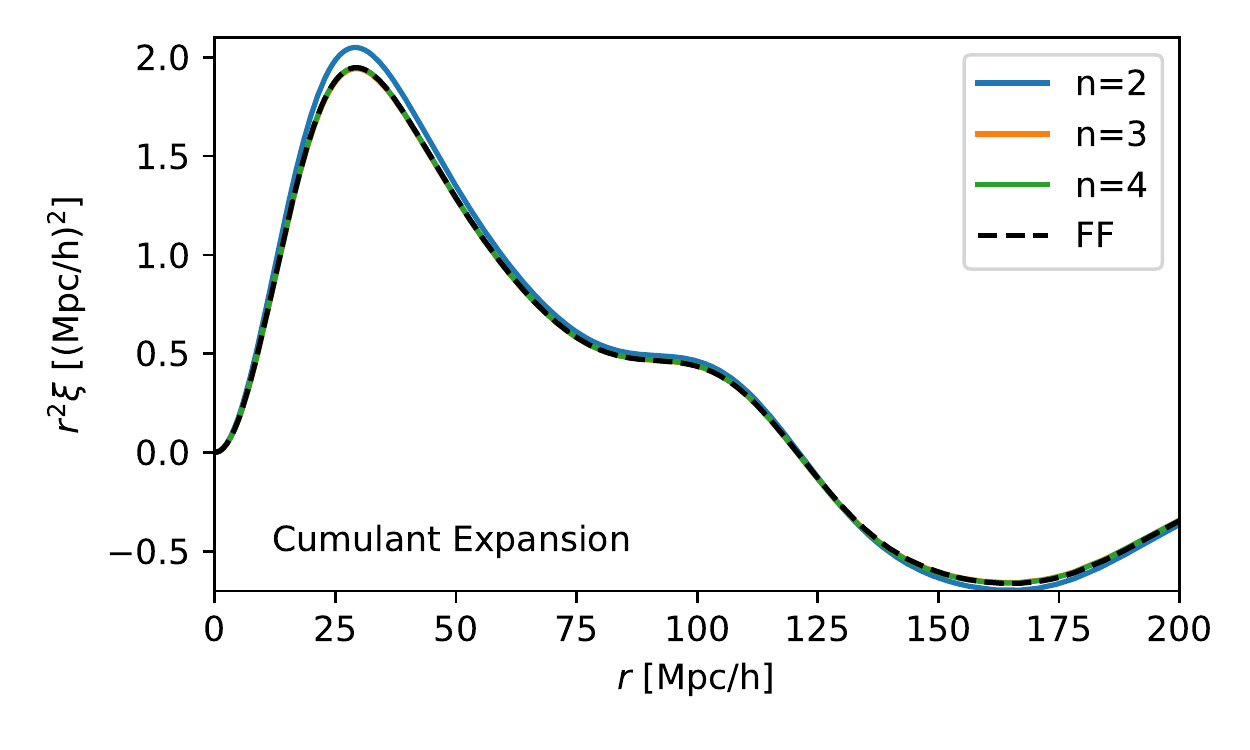}}
\caption{Series convergence for the flux two-point function in our idealized simulations with $R = 10\,h^{-1}$Mpc using the moment (left) and cumulant (right) expansions. The cumulant expansion (Eq.~\ref{eqn:cumulant_expansion}) converges much more rapidly than the moment expansion (Eq.~\ref{eqn:moment_expansion}).}
\label{fig:series10}
\end{figure}

\begin{figure}
\resizebox{0.48\textwidth}{!}{\includegraphics{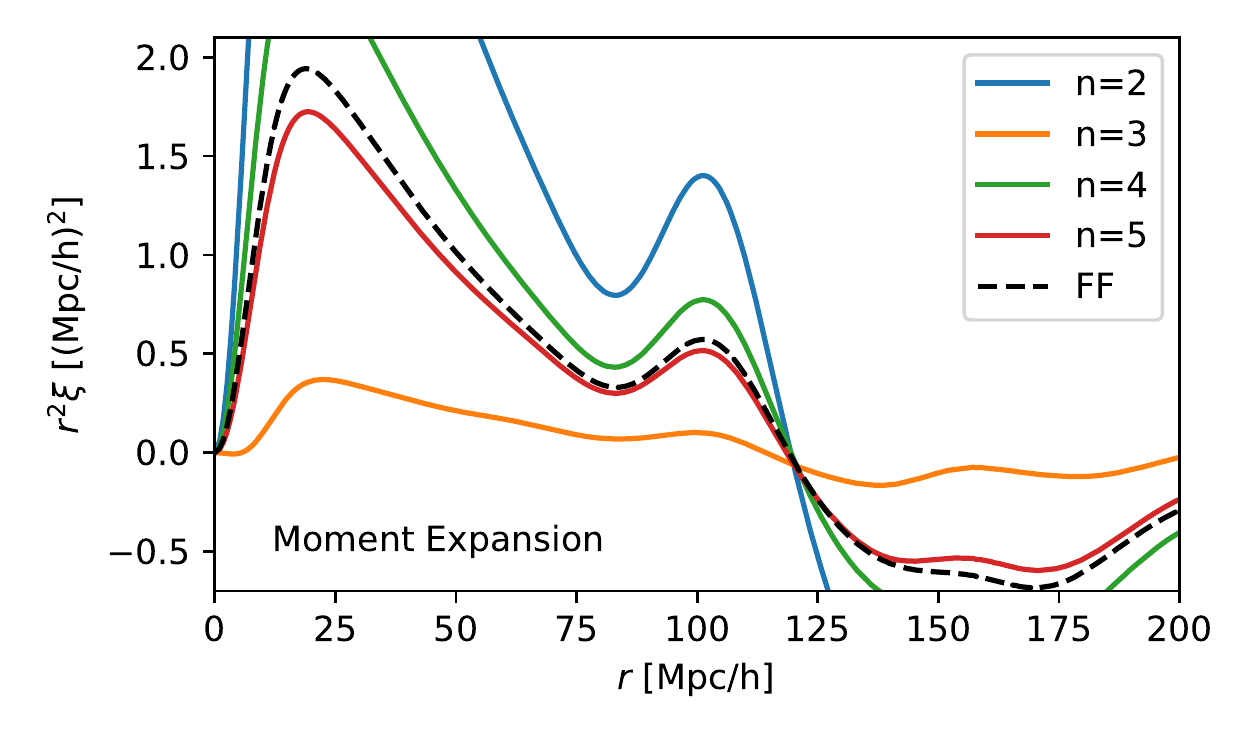}}
\resizebox{0.48\textwidth}{!}{\includegraphics{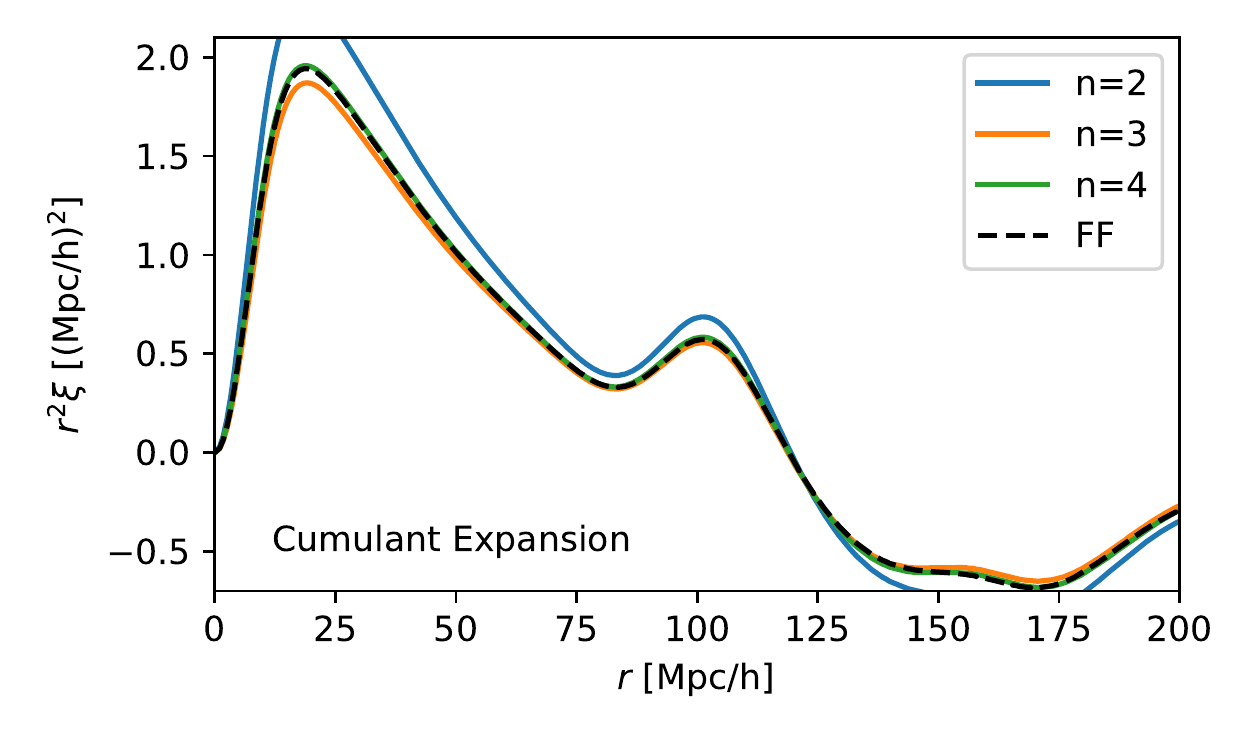}}
\caption{As for Fig.\ \ref{fig:series10} but with $R = 5\,h^{-1}$Mpc. The cumulant expansion converges much more rapidly than the moment expansion, though not as rapidly as for $10\,h^{-1}$Mpc smoothing (Fig.\ \ref{fig:series10}).  However note that the nature of the convergence is different than might have been expected, since large scales don't converge more quickly than small scales (see text for further discussion).}
\label{fig:series05}
\end{figure}

Figures \ref{fig:series10} and \ref{fig:series05} compare the moment and cumulant expansions to the direct measurement of the pseudo-flux correlation function, $\xi_{FF}$.  In each case the left hand panel shows the expansion in moments:
\begin{equation}
    F_0^2 (1 + \xi_{FF}) = \avg{e^{-\tau_1 - \tau_2}} = \sum_{n=0}^{\infty} \frac{(-1)^n}{n!} \avg{(\tau_1+\tau_2)^n}
\end{equation}
truncated at $n=2$, 3, 4 and 5. In order to minimize the difference due to the ``1'' piece on the left-hand side, we have subtracted off the contributions from the disconnected contributions summing up to the mean flux squared, i.e.
\begin{equation}
    F_0^2 \xi_{FF} =  \sum_{n=0}^{\infty} \frac{(-1)^n}{n!} \Big( \avg{(\tau_1+\tau_2)^n} - \sum_{m=0}^n \binom{n}{m} \avg{\tau^m}\avg{\tau^{n-m}} \Big),
\label{eqn:moment_expansion}
\end{equation}
where we have used $F_0^2 = \sum_{n=0}^\infty (-1)^n \avg{\tau^n}/n!.$ 

The right panel shows the same correlators resummed via the cumulant expansion. Specifically,
we have that
\begin{equation}
    1 + \xi_{FF} = \exp\Bigg\{ \sum_{n=0}^\infty \frac{(-1)^n}{n!} \big(\avg{(\tau_1+\tau_2)^n}_c - 2 \avg{\tau^n}_c \big) \Bigg\}
\label{eqn:cumulant_expansion}
\end{equation}
where the constant $F_0^2$ piece are divided out using $\ln F_0 = \sum_{n=0}^\infty (-1)^n\avg{\tau^n}_c/n!$. This is equivalent to Equation~\ref{eqn:main}. Note that, comparing to the moment expansion above, there is no explicit dependence on the mean flux $F_0^2$ in the cumulant expression.

The low order moments provide quite poor approximations to the measured flux correlations.  By contrast the cumulant expansion, Eq.~(\ref{eqn:cumulant_expansion}), shown in the right hand panel converges very rapidly, at least for our artificially large smoothing scales.  The difference between the two expansions, even at $n=2$, might be somewhat surprising given that the two-point function of the optical depth $\avg{\delta\tau_1 \delta \tau_2}$ is much smaller than unity; indeed the difference between the blue curves in the right and left panels comes mostly from the normalization factors at second order
\begin{equation}
    \xi^{n=2}_{FF} = \frac{\tau_0^2}{F_0^2} \avg{\delta \tau_1 \delta \tau_2} \text{ (moment) and } \xi^{n=2}_{FF} \approx \tau_0^2 \avg{\delta \tau_1 \delta \tau_2} \text{ (cumulant)},
\end{equation}
where in the second relation we have used $|\xi_{11}| \ll 1$. While these two expressions are equal order-by-order in $\tau$, the convergence of $F_0$ is slow.

At this point it is worth pointing out a particularly interesting feature of the convergence of the cumulant expansion, most easily seen in the right panel of Figure \ref{fig:series05}.  From long experience with Taylor series, or intuition built from cosmological perturbation theory for the matter or galaxy power spectrum, we are used to series converging more or less rapidly as a function of scale.  For example linear perturbation theory suffices to model the matter power spectrum at very low $k$, then 1-loop improves this to intermediate $k$ and so on.  We see a different pattern in Figure \ref{fig:series05}: the $n=2$ expression doesn't match $\xi_{FF}$ even above $150\,h^{-1}$Mpc, including $n=3$ improves this across all scales but there is still disagreement at both large and small scales (e.g.\ $25\,h^{-1}$Mpc and $160\,h^{-1}$Mpc), then $n=4$ improves the agreement even further and so on.  This raises an interesting question about why a linear theory form matches Ly$\alpha$ forest data at large scales, that we will discuss fruther in the next section.

Our numerical experiments validate our cumulant expansion and give some insight into the behavior of the individual terms.  If we were to continue to reduce the pressure smoothing scale to better approximate our Universe we expect to see the higher cumulants come to play an increasingly important role (e.g.\ by generating an order-unity velocity bias effect).  To understand the implications of this for measurements of the flux correlation function we need to understand the general structure of these cumulants, which we turn to in the next section.

\section{Structure of the theory}
\label{sec:structure}

Key to understanding the behavior of the flux correlation function is that of the correlators, $\xi_\tau^{(mn)}$.  In this section we first show that for Gaussian $\tau$ the flux correlation function becomes a series in $\xi_\tau^{(11)}$.  For the more general, and relevant, case where $\tau$ is non-Gaussian the behavior is more complex and more interesting.  In particular we find that when either non-linear dynamics or non-linear bias causes $\tau$ to be non-Gaussian --- even when the linear $\delta$ is Gaussian --- the higher order correlators ($\xi_\tau^{(mn)}$ for $m$ or $n$ greater than 1) can ``renormalize'' the lowest order correlator $\xi_\tau^{(11)}$ at large scales.  This behavior will be key to understanding the large-scale behavior of the flux correlation function and power spectrum.  We present this calculation first in a toy model, the peak-background split, and then motivate the behavior with perturbation theory.

\subsection{Gaussian limit}

Our Eq.~(\ref{eqn:main}) becomes particularly simple in the limit that $\delta\tau$ is a Gaussian field, since in that case only $\xi^{(11)}_{\tau}$ is non-zero.  For Gaussian $\delta\tau$ therefore
\begin{equation}
    1 + \xi_{FF} = \exp\left[ \tau_0^2 \xi^{(11)}_\tau \right]
    \quad\Rightarrow\quad
    \xi_{FF} = \sum_{n=1}^\infty \frac{\tau_0^{2n}}{n!}\left( \xi^{(11)}_\tau \right)^n
\end{equation}
a result that was previously derived by ref.~\cite{Szalay88} and can be proved directly using Price's theorem \cite{Price58}.  If $\xi^{(11)}_\tau$ drops rapidly at large scales we see that the flux correlation function is simply proportional to the linear theory correlation function on large scales (see below) while at smaller scales higher powers of $\xi^{(11)}_\tau$ introduce a scale dependent bias.  While this simple example neatly illustrates how the exponential map (going from $\tau$ to $F$) generates non-linearity, this model is not very relevant to the Ly$\alpha$ forest so we turn our attention to more complex cases.

\subsection{Peak-background split}
\label{sec:PBS}

Let us consider a simple toy model that illustrates the beyond-Gaussian and non-linear behavior.  We will employ the peak-background split (PBS), first at the level of the density field, $\delta$, and finally at the level of the optical depth, $\delta\tau$.  To start we will neglect redshift-space distortions, deferring them until we discuss perturbation theory.

First we approximate the density field as a sum of long- and short-wavelength modes: $\delta=\delta_\ell + \delta_s + \nu_2\delta_\ell\delta_s$ with $\nu_2=34/21$ the angle average of the Eulerian perturbation theory kernel, $F_2$ \cite{Ber02}. The $\nu_2$ term approximates the effects of non-linear mode coupling in this model.  We consider the case where $\delta_s$ is uncorrelated on the separation scale, $\mathbf{r}$, of interest and $\delta_\ell$ is a Gaussian random variable (i.e.\ a ``linear'' mode). For an optical depth $\tau[\delta]$ that is an arbitrary local function of the nonlinear density we can write to first order in the long-wavelength density
\begin{equation}
    \tau[\delta] = \tau[\delta_s + (1 + \nu_2 \delta_s) \delta_\ell] = \tau[\delta_s] + \tau'[\delta_s] (1 + \nu_2 \delta_s) \delta_\ell + \mathcal{O}(\delta_\ell^2).
    \label{eqn:pbs_seljak}
\end{equation}
suggesting that the (unnormalized) linear bias is $b_\tau = \avg{\tau' + \nu_2 \delta \ \tau'}_s$ where primes stand for derivatives with respect to the argument and the average is carried out in the absence of long modes. This recovers the results of ref.~\cite{Seljak12}, and indeed carries over without modification to the linear flux bias in this limit as well given that it is just given by the composite map $F[\delta] = \exp(-\tau[\delta])$. This ease of translation does not carry over to redshift space however, as we will see below.

Let us now treat the specific case where $\tau[\delta]$ is given by a bias expansion. This will allow us to anticipate qualitatively what happens within the full perturbation theory argument. Recalling the $\delta_s$ are uncorrelated at separation $\mathbf{r}$, the lowest order cumulant in real-space is
\begin{equation}
    \xi_\tau^{(11)} = \left\langle \delta\tau_1\delta\tau_2 \right\rangle
    = b_1^2\langle\delta_{\ell 1}\delta_{\ell 2}\rangle
    + \frac{b_1b_2}{2} \left\langle \delta_1\,\delta_2^2 \right\rangle
    + \cdots
\label{eqn:tau11}
\end{equation}
where $\delta\tau_j=\delta\tau(\mathbf{r}_j)$ and similarly for $\delta$.
While the higher order correlators of the purely Gaussian piece, $b_1\delta_{\ell}$, vanish,  the higher order bias and mode coupling terms in the higher cumulants generate contributions like $\langle\delta_{\ell 1}\delta_{\ell 2}\rangle$.  Consider the $b_1b_2$ term above.  Since $\delta_2^2$ contains $2\delta_{s2}\ \nu_2\delta_{\ell 2}\delta_{s2}$ we have
\begin{align}
    \left\langle \delta_1\,\delta_2^2 \right\rangle
    \supset 2\nu_2\left\langle  \delta_{\ell 1} \delta_{s2} \delta_{\ell 2}\delta_{s2} \right\rangle
    = 2\nu_2\left\langle\delta_{s}^2\right\rangle\ \xi_L
\label{eqn:xitau11_pbs}
\end{align}
with $\xi_L\equiv \langle\delta_{\ell 1}\delta_{\ell 2}\rangle$ the long-mode density correlation function.  This contribution depends upon small-scale physics through $\langle\delta_s^2\rangle$ and ``renormalizes'' the lowest order expression $b_1^2\xi_L$ \cite{McDRoy09}.  We will see a direct analog of this term in the next section.  Another such term is
\begin{equation}
    \xi_\tau^{(12)} = \left\langle\delta\tau_1\,\delta\tau_2^2\right\rangle
    \supset 2b_1^3\nu_2\left\langle (\delta_{\ell 1}+\delta_{s 1})\,(\delta_{\ell 2}+\delta_{s2})\delta_{\ell 2}\delta_{s2} \right\rangle
    \supset 2b_1^3\nu_2\left\langle\delta_s^2\right\rangle\ \xi_L
\label{eqn:xitau12_pbs}
\end{equation}
which also depends upon small-scale physics and ``renormalizes'' the lowest order term.  Again, there will be a direct analog in the next section.
In fact there are an infinite number of such terms that arise in the cumulant expansion.

\begin{figure}
    \centering
    \includegraphics[width=0.9\textwidth]{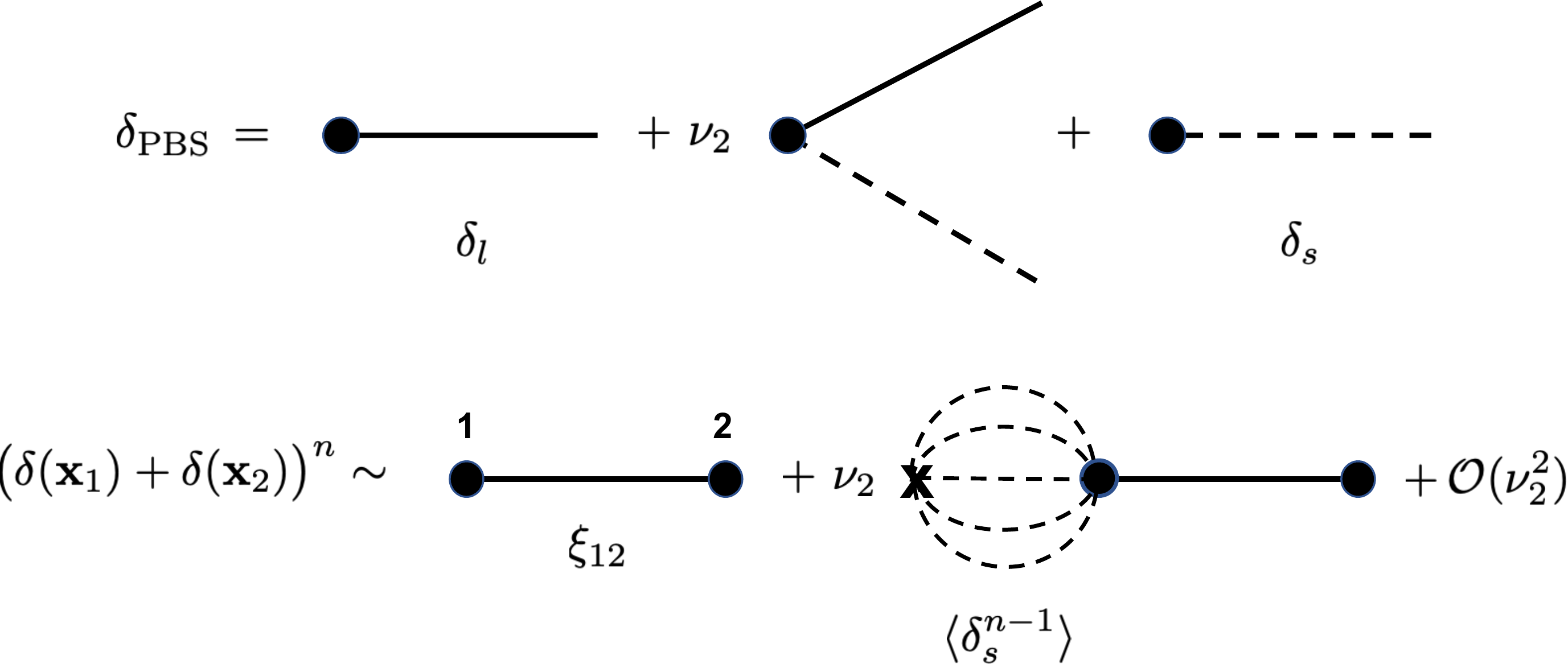}
    \caption{(Top) Contributions to the nonlinear density in the peak-background split. Solid points indicate points in configuration space, while the solid and dashed lines refer to long and short modes which are coupled with strength $\nu_2$ in the middle diagram. (Bottom) Diagrams indicating the terms that contribute to the flux correlation function in the PBS approximation at order $n$ in the optical depth. While the long modes are Gaussian, the short modes can couple to itself at a point to arbitrary order, though not at other points. We have omitted a similar diagram proportional to $\nu_2^2$ that has a corresponding short-mode ``bubble'' on point on the right.}
    \label{fig:PBS}
\end{figure}

We can see the manner in which the higher order terms combine in the two-point function calculation by simplifying our peak-background split model even further.  Let us focus on the case where the optical depth is some constant multiple of the nonlinear density, i.e.\ $\tau = \tau_0 (1 + \delta(\bx))$, with the density expressed via the PBS, as this form illustrates our point quite clearly.  In this limit the correlation function is simply
\begin{equation}
    \avg{F(\bx_1) F(\bx_2)} = e^{-2\tau_0} \sum_{n=0}^\infty \frac{(-\tau_0)^n}{n!} \Bigg( \sum_{i=1,2}  \big[ \delta_\ell + (1 + \nu_2 \delta_\ell) \delta_s \big](\bx_i) \Bigg)^n.
    \label{eqn:corr_pbs}
\end{equation}
Our assumption that the long and short modes are uncorrelated, and that the former is Gaussian, then leads to three kinds of contributions to the correlation function, shown diagramatically in Figure \ref{fig:PBS}. First, the long modes can be contracted between the two points in the second cumulant. Then, the contributions proportional to $\nu_2$ can couple either with the pure long-mode or itself. Finally, the pure short-mode contributions must be contracted at a single point and, along with the point contractions of the long modes, amount to the normalization, $F_0^2$. From these considerations we can extract terms directly proportional to the linear theory correlation function in the cumulants at order $n$:
\begin{equation*}
    2 n (n-1) \nu_2 \avg{\delta_s^{n-1}} \xi_{L} + n (n-1) \nu_2^2 \sum_{m=0}^{n-2} \avg{\delta_s^{m+1}} \avg{\delta_s^{n-2-m+1}} \xi_{L}
\end{equation*}
where the combinatorial factors denote the number of possible contractions of long and short modes. In particular, we will see in the next section that the $n=3$ term exactly corresponds to the UV-sensitive piece in perturbation theory for real space, with $\nu_2$ standing in for the matter density kernel $F_2$ and the bubble with $\avg{\delta_s^2}$ standing in for the large momentum modes in the integral. Summing up all the contributions and normalizing by the total flux yields 
\begin{equation}
    F_0^2\, \xi_{FF} = \left( \vphantom{\int}
    \left\langle F'[\delta] \right\rangle_{s} + \nu_2 \left\langle \delta\, F'[\delta] \right\rangle_{s} \right)^2 \xi_{L}
    + \mathcal{O}(\delta_\ell^4)  .
\label{eqn:pbs_xi_ff}
\end{equation}
This agrees with the field-level derivation in Eq.~\ref{eqn:pbs_seljak}, as well as similar results in refs.~\cite{Seljak12,Arinyo15,Cieplak16}. 

Let us close by noting the limitations of the above calculations. Firstly, the averages over the short modes (e.g.\ $\avg{\delta^2}_s$) depend on small scale physics like the suppression of density power below the Jeans scale due to gas pressure and additional highly nonlinear astrophysical processes. Moreover, the PBS calculations above assume that the nonlinearities involved can be modeled as a single density coupling between long and short modes, given by a coefficient $\nu_2$ fixed by second-order perturbation theory when there is in fact no restriction on higher-order couplings or those involving velocities, which are themselves subject to different small scale nonlinearities (e.g.\ turbulence). Despite these complications, however, the fact that on large scales the theory prediction boils down to bias coefficients multiplying the linear correlation function is significant, showing that the small scales can be modeled via a handful of effective-theory parameters even without a full modelling of small scale astrophysics.

\subsection{One-loop perturbation theory}
\label{sec:one_loop}

In the previous section we looked at how the cumulants behaved in a simplified model. In this section we will show that similar behavior occurs at 1-loop order in perturbation theory, where the proper inclusion of redshift-space distortions also gives rise to the phenomenon of velocity bias. In redshift space we assume that the optical depth fluctuation is determined by shifting the real space fluid element\footnote{See also ref. \cite{Cieplak16} for some further discussion on the validity of this mapping for Ly$\alpha$ forests and the effects of thermal broadening.}
\eeq{
1 + \df \tau_s(\vec r) 
= \int d^3 \tilde{r}~ \lb 1 + \df \tau (\tilde{\vec r}) \rb 
\df^D\lb \vec r - \tilde{\vec r} - \vec u_\pp \hat n \rb,
}
where $\vec u_\pp$ is the tracer velocity component projected along the line of sight.
This is equivalent to the galaxy overdensity redshift space mapping and we are thus able to use the same Eulerian perturbative kernels $Z_n$ (see e.g. \cite{Ber02}) 
to describe the perturbative orders of $\df \tau_s$ field.

Compared to the standard 2-point function calculation for biased tracers in redshift space, the Ly$\alpha$ calculation at 1-loop involves only one additional cumulant. This is because at 1-loop order cumulants with more than three powers of the optical depth $\delta \tau$ vanish since, for example, the fourth moment $\avg{\delta \tau \delta \tau \delta \tau \delta \tau} \sim \avg{\delta \tau\delta \tau} \avg{\delta \tau\delta \tau}$ is given at leading order by disconnected contributions. Starting from the Eq.\eqref{eqn:main} which constitutes 
the cumulant expansion for the flux correlators, then,
we have
\eeq{
\ln \Big( 1 + \la \delta F \delta F' \ra_s \Big)
= \tau_0^2  \xi_s^{(11)} (\vec s) - \tau_0^3  \xi_s^{(12)} (\vec s) + \mathcal{O}(P_L^3).
\label{eqn:flux_xi_1loop}
}
Exponentiating, expanding and keeping only the 
one-loop results we obtain the equivalent expression for the 
flux power spectrum
\eeq{
  P_{FF} (\vec k) 
  = \tau_0^2 P_{s}^{(11)} (\vec k)
  - \tau_0^3 P_{s}^{(12)} (\vec k)
  + \frac{1}{2}\tau_0^4 \int_{\vec p} 
  P_{s}^{(11)} (\vec p) 
  P_{s}^{(11)} (\vec k - \vec p)
  + \cdots ,
\label{eqn:flux_PS_1loop}
}
where we have introduced the cumulant spectra
\eeq{
P_{s}^{(11)} (\vec k) = \int d^3 s ~\xi_s^{(11)} (\vec s) e^{i \vec k \cdot \vec s}
\quad{\rm and}\quad
P_{s}^{(12)} (\vec k) = \int d^3 s ~\xi_s^{(12)} (\vec s) e^{i \vec k \cdot \vec s},
}
and the last term is obtained as the 
Fourier transform of the second term in 
the expansion of the first cumulant. 
In order for this term to be 
consistently evaluated in one-loop perturbation 
theory only linear level (Kaiser) contributions 
to $P_{s}^{(11)}$ are required, 
besides the appropriate counterterms that 
are also discussed in Appendix \ref{app:ept_calc}.

The $(11)$ cumulant is simply the redshift-space two-point function for biased tracers, which is well known.  Let us focus on the cumulant $\xi_s^{(12)}$. There are two contributions to this cumulant at 1-loop:
\begin{equation*}
    \avg{\delta \tau_1 \delta \tau_2^2} = 2\avg{\delta \tau^{(1)}_1 \delta \tau^{(1)}_2 \delta\tau^{(2)}_2} + \avg{\delta \tau^{(2)}_1 \delta \tau^{(1)}_2 \delta\tau^{(1)}_2}.
\end{equation*}
These are
\begin{equation}
    P^{(12)}_{s} \supset
    2 \times 2 Z_1(\bk) P(k)\int\frac{d^3\bp}{(2\pi)^3}\ Z_1(\bp) Z_2(\mathbf{k},\bp)P(p).
\label{eqn:xi112_main}
\end{equation}
and
\begin{equation}
    P^{(12)}_{s} \supset 2 \int\ \frac{d^3 p}{(2\pi)^3}\ Z_2(p, k-p) Z_1(-p) Z_1(p-k) P(p) P(k-p).
\label{eqn:xi112_other}
\end{equation}
The $Z_n$ are the order $n$ perturbation theory kernels in redshift space (\cite{Ber02} and Appendix \ref{app:ept_calc}) with $Z_1(\bk_1) = b_1 + f\mu_1^2$ and
\begin{align}
    Z_2(\bk_1, \bk_2) &= b_1F_2(\bk_1,\bk_2) + f\mu_k^2 G_2(\bk_1,\bk_2) + \cdots \nonumber \\
    &+ \frac{fk\mu_k}{2} \left[ \frac{\mu_1}{k_1}(b_1+f\mu_2^2) + \frac{\mu_2}{k_2}(b_1+f\mu_1^2) \right]
\end{align}
where $\mu_k = \hk \cdot \hn$, $\bk = \bk_1 + \bk_2$ and $F_2$ and $G_2$ are the standard density and velocity kernels in Eulerian perturbation theory \cite{Ber02}.  (In the definitions above we have omitted the contributions from higher-order bias such as $b_2$ and $b_s$ for the sake of brevity. These do not qualitatively affect our conclusions and the full calculation with all relevant terms included are detailed in Appendix \ref{app:ept_calc}.)

The integral in Equation~\ref{eqn:xi112_main} has a form where the linear power spectrum is multiplied by a scale-dependent term that is sensitive to contributions from small-scale modes $p \gg k$. In the large-scale ($k \rightarrow 0$) limit we can write
\begin{equation}
    Z_1 Z_2 \supset \left(b_1 + f (\hat{p} \cdot \hn)^2 \right)\Bigg[ \left(\frac{5}{7} + \frac{2}{7}\ [\hp \cdot \hk]^2 \right) b_1 + \left(\frac{3}{7} + \frac{4}{7}[\hp \cdot \hk]^2 \right) f \mu_p^2 + \frac{f \mu_p^2}{2} \left(b_1 + f(\hk \cdot \hn)^2  \vphantom{\int} \right) \Bigg]\nonumber.
\end{equation}
The dependence on the long-wavelength wavevector $\hk$ can be pulled out of the above integrand to give (see Appendix \ref{app:ept_calc})
\begin{equation}
    P^{(12)}_{s}(\bk) \rightarrow 2 (b_1 + f\mu_k^2) (A_0 + A_2 \mu_k^2)\ \sigma^2P_L(k)
\end{equation}
where $\sigma^2 = \int_{\vec p} P_L(p)$ is the rms density contrast and $A_{0,2}$ are functions of the bias parameters and growth rate given in the Appendix \ref{app:ept_calc}. The above form strongly suggests a modification to the Kaiser form of redshift-space distortions
\begin{equation*}
    \big[ (b_1 + \Delta b_1) + (f + \Delta b_\mu)\mu_k^2 \big]^2 P_L(k) = (b_1 + f\mu_k^2)^2 P_L(k) + 2 (b_1 + f\mu_k^2) (\Delta b_1 + \Delta b_\mu \mu_k^2) P_L(k) + ...
\end{equation*}
However, it is important to note that the small-scale contributions $\Delta b_{1,\mu}$ need to have their UV sensitivities renormalized by appropriate counterterms, so that their value cannot be determined by the 1-loop calculation above.

Before moving on to the most general case let us comment briefly on some features of the above calculation. Most importantly, the inclusion of redshift-space distortions produces a novel feature, velocity bias, wherein the linear-theory prediction for biased tracers is modified when applied to the Ly$\alpha$ flux. Previous studies \cite{Seljak12,Cieplak16} of this effect have largely focused on the effect of dynamical nonlinearities in the density; indeed, in the absence of RSD (i.e.\ $f=0$) our derivation recovers the PBS result in Equation \ref{eqn:pbs_xi_ff} via the angular average $\nu_2 = 2 \avg{5/7 + 2(\hat{p}\cdot\hat{k})^2/7}$. A new result in this paper is that in redshift space contributions coupling the orientation $\hk$ of the long mode and the short modes contribute at equal order owing to both the nonlinear velocity kernel ($G_2$) and displacements due to the real-to-redshift-space mapping. Finally, a salient feature of our derivation is that the $\hk$ dependence of the integrand can be written as
\begin{equation*}
    \int d\Omega_{\bp}\ Z_1(\bp) Z_2(\bp,\bk)\; \rightarrow a^{(0)} + a^{(2)}_{ij} \hk_i \hk_j
\end{equation*}
where the tensor $a^{(2)}$ is expressible in terms of $\delta_{ij}$ and $\hn_i$ due to the axial symmetry of redshift-space distortions. Thus while the coefficients themselves depend on complicated small-scale physics the parametric contribution is only quadratic in $\mu^2$.

Lastly let us look at the stochastic contributions and Equation~\ref{eqn:xi112_other}. As we have mentioned, the structure of the $P^{(11)}_{s}$ cumulant is equivalent to the galaxy case, see e.g.\ ref.~\cite{Chen20} for expressions in the same bias convention including a detailed discussion of stochastic contributions. Equation~\ref{eqn:xi112_other} generates the same functional form for stochastic contributions in the second cumulant in the large-scale limit (Appendix~\ref{app:ept_calc})
\eeq{
P^{(12)}_{s} \supset
{\rm ``const_0"} 
+
{\rm ``const^{(0)}_2"} \frac{k^2}{k_\star^2}
+ {\rm ``const^{(2)}_2"} \mu^2 \frac{k^2}{k_\star^2}
+ \ldots,
}
which consequently renormalizes the contributions given in $P^{(11)}_{s}$.
However, note that the higher cumulants introduce an explicit dependence on the logarithmic growth rate $f$ to the counterterms ${\rm ``const_0"}$ and ${\rm ``const_2"}$. Importantly, the scale independent, white noise contribution remains isotropic.

\subsection{General structure of the large-scale limit}
\label{sec:more_loops}

\begin{figure}
    \centering
    \includegraphics[width=\textwidth]{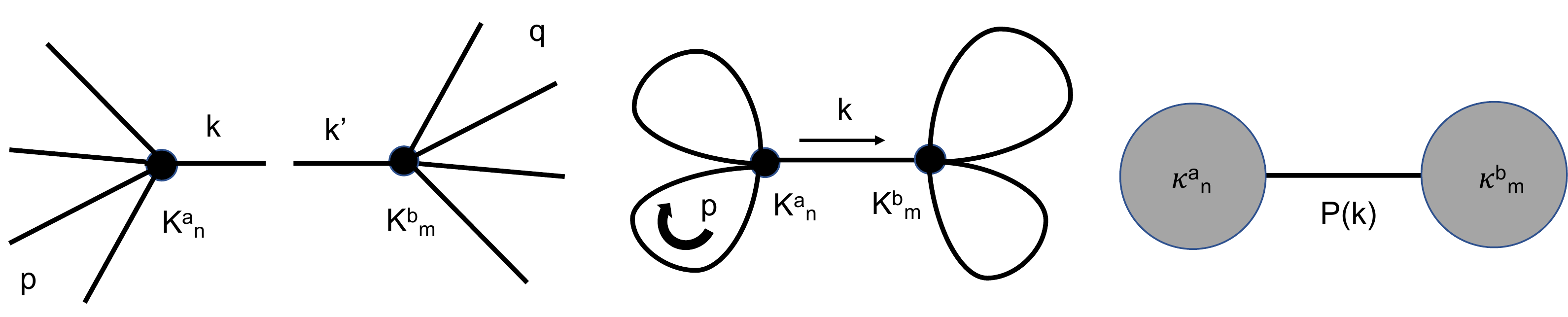}
    \caption{Diagramatic structure of nonlinearities contributing linear-theory-like terms. These consist of pairs of $n^{\rm th}$ order vertices with one long-wavelength leg ($\bk, \bk'$), with the other small-scale modes ($\bp,\bq$) integrated out at each point, leading to a factorizable structure where nonlinear bubbles dress linear theory at each point.
    }
    \label{fig:beyond_one_loop}
\end{figure}

While we have shown it explicitly for the 1-loop spectrum, the large-scale angular dependence (the square of a quadratic in $\mu$) holds beyond 1-loop.  This can be established by looking at how the mode-coupling kernels behave in the long-wavelength limit.  The contribution going as $P(k)$ as $k\to 0$ can be isolated by considering diagrams linking one power of the linear field at each point with all other momenta contracted with themselves to give loops at each of the two vertices (see Fig.~\ref{fig:beyond_one_loop}).  Defining
\begin{equation}
    \kappa^{(a)}_n(\bk) =  \int_{p_i}\ K^{(a)}_n(k, p_1, ..., p_{n-1}) \avg{\delta_0(p_1) ... \delta_0(p_{n-1})\ (2\pi)^3 \delta^D(\sum p_i)},
\end{equation}
as the integral over the $(n-1)/2$ short-wavelength loops at one vertex that comes from $\delta\tau^a$ at $n^{\rm th}$ order in the linear density (the last mode being the $\delta(\bk)$ to be contracted with the other vertex) we see as $k\to 0$ the diagram in Fig.~\ref{fig:beyond_one_loop} becomes
\begin{equation}
    \sum_{a,b}\sum_{n,m} \kappa_n^{(a)}\kappa_m^{(b)} P_L(k) = \left( \sum_a \sum_n \kappa_n^{(a)} \right)^2 P_L(k)
    \rightarrow b^2(\mu) P_L(k) \quad (k \rightarrow 0)
    \label{eqn:kappas}
\end{equation}
where $b(\mu)$ is the large-scale asymptote of the sum of $\kappa^{(a)}_n$'s which will tend to have non-constant $k$ dependence outside of that limit. That we can factor these diagrams simply into the square of an angular-dependent bias factor $b(\mu)$ is a result of the independence of the loops on each side of Fig.~\ref{fig:beyond_one_loop}. This also directly implies that cross correlations with the Ly$\alpha$ forest can be parametrized in the usual way, i.e. 
\begin{equation*}
    P^{\rm g,Ly\alpha}_s(\bk) = (b^{\rm g}_1 + f\mu^2)\ b^{\rm Ly\alpha}(\mu) P_L(k)
\end{equation*}
at linear order.  This raises the possibility of improving constraints on the growth rate through ``sample variance cancellation'' \cite{Seljak09} with biased tracers like quasars that obey the Kaiser form even if Ly$\alpha$ does not\footnote{Indeed, soon after this paper was submitted \cite{Cuceu21} reported promising results on high-redshift growth rate measurements via Ly$\alpha$-quasar cross correlations using this technique.}.   Even though Ly$\alpha$ does not have a simple dependence on the growth rate, $f(z)$, as we will now show the functional dependence of $b(\mu)$ is itself limited to two free parameters which can be leveraged in the modeling. Futhermore, we note that at the 1-loop level $\kappa^{(a)}_n$ can also be used to capture all $P^{(13)}$ type contributions by expanding $\kappa^{(a)}_n \supset \alpha_0 k^2 + \alpha_2 k^2 \mu^2 + \cdots$ beyond zeroth order in the wavevector.

To complete the picture we need to show that $b(\mu)$ is at most quadratic in $\mu$. As a first step we note that, as was the case in the 1-loop case (Eq.~\ref{eqn:xi112_main}), the $n^{\rm th}$ order kernel contributing to $\delta\tau^a$, $K^{(a)}_n$, can be written as a product of redshift-space kernels $Z_n$, only one of which contains the long mode $\bk$. The problem can then be reduced to the angular structure of the redshift-space kernels alone, which indeed must truncate at $\hk_i \hk_j$ as we show in Appendix \ref{app:general_structure_k_to_0} as a consequence of the structure of dynamical nonlinearities in structure formation. This implies that $b(\mu)$ must be quadratic in $\mu$ to all orders in perturbation theory.

Let us close with some comments on how this argument may be extended to operators beyond linear bias. As we have shown above, modifications to the Kaiser formula arising from short modes contracted at a point lead to a factorable but anisotropic bias term. We can extend this argument to the case of multiple long modes
\begin{equation}
    \kappa^{(a)}_n(\bk_1, ..., \bk_m) =  \int_{p_i}\ K^{(a)}_n(k_1, ... k_m, p_1, ..., p_{n-m}) \avg{\delta_0(p_1) ... \delta_0(p_{n-m})\ (2\pi)^3 \delta^D(\sum p_i)}.
\end{equation}
Such contributions will need to be renormalized with their own (anisotropic) counterterms which will again be limited in their forms by the structure of dynamical nonlinearities in the equations of motion (i.e.\ $Z_n$). Recently, refs.~\cite{Givans20,Des18} proposed a second-order anisotropic bias basis in the context of Ly$\alpha$ and a third-order one for generic biased tracers with line-of-sight selection effects. In Appendix \ref{app:general_structure_k_to_0}, we briefly sketch the kinds of terms that are generated from the procedure outlined above, showing that they appear to be well-parametrized by the bases proposed in those works.  We intend to return to a more systematic study of these terms and further constraints on their forms due to fundamental symmetries \cite{Fujita+:2020} in a future work.

\section{Conclusions}
\label{sec:conclusions}

Modern surveys capable of measuring the spectra of hundreds of thousands or millions of distant objects with modest signal to noise per \AA\ can tightly constrain the flux decrement power spectrum on cosmological scales \cite{McQuinn11a,Font18}.  In turn this provides tight constraints on the distance-redshift relation \cite{Bourboux20} and provides a new method for measuring the clustering of high redshift objects \cite{Font12}.

The transmitted flux in the Ly$\alpha$ forest probes the underlying density field in a non-linear manner, but on large scales the flux power spectrum is proportional to that predicted in linear perturbation theory \cite{Bourboux20}.  In this paper we have investigated how large- and small-scale physics affect the measured power spectrum by expanding the flux correlation function in cumulants of the optical depth field (Eq.~\ref{eqn:main}).  Since the optical depth can be related to the density field by a bias expansion (Eq.~\ref{eqn:bias_expansion}), this provides a direct connection between the large-scale, cosmological physics of interest and the measurement.  Our formalism gives a systematic means of computing corrections to the flux power spectrum arising from non-linearities, redshift-space distortions and scale dependent bias.  It provides an easy way to see the emergence of ``velocity bias'' in the measured Ly$\alpha$ forest power spectrum in redshift space and the manner in which small-scale physics (i.e.\ ``astrophysics'') affects the prediction.

We demonstrated in some idealized numerical experiments (\S\ref{sec:numerics}) that the convergence of the cumulant expansion does not behave as one might expect from experience with the matter power spectrum in cosmological perturbation theory.  Higher cumulants affect the flux correlation function at both large and small scales, rather than being confined to increasingly small scales.

To understand the way in which non-linearities, redshift-space distortions and scale dependent bias affect the flux correlation function, we have presented a perturbative analysis of the flux two-point function due to nonlinearities in the exponential mapping within the peak-background split (\S\ref{sec:PBS}), 1-loop perturbation theory (\S\ref{sec:one_loop}), and general considerations of the structure of the equations of motion (\S\ref{sec:more_loops}).  While the clustering signal for galaxies in redshift space has an ``unbiased'' contribution due to cosmic velocities given by the Kaiser form $(b_1 + f\mu^2)^2 P_L(k)$, it is well known \cite{McDonald03,Slosar09,Slosar11,Bourboux20} that the Ly$\alpha$ forest breaks this simple form. Within the framework of effective perturbation theory this is due to the presence of contact terms from higher cumulants which modify the large-scale limit of the power spectrum away from the Kaiser form but preserves the ``square of a quadratic in $\mu$'' behavior. While previous works have shown this within the context of density nonlinearities and the peak background split, we show using a 1-loop calculation that nonlinear contributions to the velocity, as well as displacements due to redshift-space distortions, contribute at equal order. Extending from this example we show that the quadratic-in-$\mu$ form is preserved to all orders in perturbation theory based on the structure of the equations of motion alone and outline how higher-order bias terms can be similarly generated, making connection with the anisotropic bases of refs.~\cite{Givans20,Des18} and commenting on the possibility for cross correlations and sample variance cancellation.  It is worth noting that this analysis of velocity bias should qualitatively extend to less tractable (non-exponential) maps of redshift-space density fields, as has been for example observed in the clustering of cosmic voids \cite{Chuang17}.

Our calculations demonstrate that for Ly$\alpha$, unlike for galaxies, both the isotropic and anisotropic contributions to the clustering depend on small-scale astrophysics.  In particular, the coefficients of both the $\mu^0$ and $\mu^2$ terms in the linear bias receive contributions from small-scale physics at arbitrary order.  While of little direct relevance for measuring distances using baryon acoustic oscillations, this fact has immediate implications for extracting physics from the broad-band shape of the Ly$\alpha$ forest flux power spectrum.  If we identify the piece of the spectrum that can be predicted perturbatively with ``cosmological information'' and the small-scale physics as ``astrophysical information'' we see that the Ly$\alpha$ forest contains an irreducibly complex mixture of the two.  At the practical level this leads to a larger number of coefficients that need to be modeled or fit to measurements, which necessarily reduces the amount of information that can be obtained.  However unlike for the case of galaxies, the Ly$\alpha$ forest can be simulated quite well which may provide a means of using informative priors on the coefficients to improve constraints. An interesting avenue of further investigation may be to measure these coefficients in simulations via the recently developed anisotropic separate universe simulation technique \cite{Masaki20,Stucker21}, as was already done approximately in \cite{McDonald03}. Such methods enable the measurment of responses of e.g.\ Ly$\alpha$ clustering to large-scale tidal fields with nontrivial angular dependence such as those considered in this work (encoded in $\kappa^{(a)}_n(\bk)$ of Eqn.~\ref{eqn:kappas}). The feasibility of accurate hydrodynamical modeling of the Ly$\alpha$ forest on small scales and the fact that the impact of these physics on the broadband shape of the correlation function on large scales can be isolated into single coefficients measurable from such simulations suggests that accurate full-shape analyses of the Ly$\alpha$ forest are possible even if simulations and perturbation theory are individually restricted to small and large scales.

Finally, let us briefly comment on the implications of our work for the Ly$\alpha$ forest BAO feature. It is well known that non-linear evolution causes a broadening of the BAO peak in the correlation function of galaxies \cite{ESW07,Crocce08,Mat08a,CLPT,SenZal15,Vlah16,Blas2016} and, similar measurements being one of the premier science goals for the Ly$\alpha$ forest, it is interesting to ask whether the BAO peak in the flux correlation function is similarly broadened.  Our formalism provides a means to answer this question, since we can treat $\delta\tau$ as a biased tracer of the density field for which the role of long-wavelength displacements responsible for the damping are well understood and then relate cumulants of $\delta\tau$ to the flux correlation function using Eq.~(\ref{eqn:main}).  What our calculations make clear, however, is that the broadening of the BAO peak in the Ly$\alpha$ forest will be different at each (higher) order in optical depth than it is for tracers such as galaxies or QSOs.  Each of the cumulants in Eq.~(\ref{eqn:main}) responds to the long wavelengths differently, and thus the broadening has the potential to be more complex than a single exponential damping factor sometimes used in the literature. We defer further investigation of this to future work.

\acknowledgments

We would like to thank Andreu Font, Vid Irsic, Pat McDonald and Uros Seljak for helpful comments on a draft of this work.
S.C.\ is supported by the National Science Foundation Graduate Research Fellowship (Grant No.~DGE 1106400) and by the UC Berkeley Theoretical Astrophysics Center Astronomy and Astrophysics Graduate Fellowship.
M.W.\ is supported by the U.S. Department of Energy and the NSF.
Z.V.\ is supported by the Kavli Foundation.
This research has made use of NASA's Astrophysics Data System and the arXiv preprint server.
This research used resources of the National Energy Research Scientific Computing Center (NERSC), a U.S.\ Department of Energy Office of Science User Facility operated under Contract No.\ DE-AC02-05CH11231.

\appendix

\section{Exponential mapping and the 2-point function}
\label{app:details}


\subsection{Flux autocorrelation}

The two point function of the flux field is given by
\eeq{
\la F(\vec x) F(\vec x') \ra_F = F_0^2 \la 1 + \la \delta F(\vec x) \delta F(\vec x') \ra_{\delta F} \rb
= \la e^{- \tau (\vec x) - \tau (\vec x') } \ra_{\tau}
}
where the subscripts on $\langle\cdots\rangle$ indicates which distribution is used in the average.  We will make use of the cumulant expansion to simplify this expression.  Defining
\begin{equation}
    Z_\tau(J) = \ln\left\langle \exp\left[ \int d\mathbf{x}\ J(\mathbf{x})\tau(\mathbf{x}) \right] \right\rangle_{\tau}
    \quad\mathrm{and}\quad
  \left\langle \tau_\mathbf{x}^m \tau_{\mathbf{x}^\prime}^{n-m}\right\rangle_{\tau,c}
  \equiv \left. \frac{\partial^n}{\partial J^m_\mathbf{x} \partial J^{n-m}_{\mathbf{x}^\prime}}
  Z_\tau \right|_{J=0}
\label{eqn:tau_cumulants}
\end{equation}
we see the cumulant is the coefficient of $J_\mathbf{x}^m J_{\mathbf{x}^\prime}^{n-m}/m!/(n-m)!$ in the expansion of $Z_\tau$.  Rewriting $1/m!/(n-m)!$ as $\binom{n}{m}/n!$ and using
\eeq{
  F_0^2 =\exp \lb 2 \sum_{n=1}^\infty \frac{(-1)^n}{n!} \left\langle \tau^n \right\rangle_c \rb
}
we have
\eeq{
  1 + \la \delta F \delta F' \ra_{\delta F}
  = \exp \Bigg[ \sum_{n=2}^\infty \frac{(-1)^n}{n!} \mathcal{C}_n \Bigg]
  \quad\mathrm{with}\quad
  \mathcal{C}_n = \sum_{m=1}^{n-1} \binom{n}{m} \la \tau^m \tau'^{n-m}\ra_{\tau, c} 
\label{eqn:F2pt}
}
where we have used the shorthand $\tau^\prime = \tau(\mathbf{x}^\prime)$.

We want to express these correlators in terms of fluctuations of optical depth $\delta \tau$.
Replacing $\tau(\mathbf{x})=\tau_0(1+\delta\tau(\mathbf{x}))$ in the definition of $Z_\tau$ it follows that
\eeq{
  \la \tau^m \tau'^{n-m}\ra_{\tau, c} 
  = \frac{d^m}{dJ^m}\frac{d^{n-m}}{dJ'^{n-m}} Z_{\tau}(J) \Big|_{J=0}
  = \frac{d^m}{dJ^m}\frac{d^{n-m}}{dJ'^{n-m}} \Big[ \tau_0\int J + Z_{\delta \tau} ( \tau_0 J ) \Big] \Big|_{J=0}.
}
Since $n\geq2$ the $\tau_0\int J$ term does not contribute and we have
$\la\tau^m\tau'^{n-m}\ra_{\tau,c} = \tau_0^n \la\delta\tau^m \delta\tau'^{n-m}\ra_{\delta \tau,c}$.
Introducing 
\eeq{
  \xi^{(ij)} (\vec r) =  \la \df \tau^i \df \tau'^j \ra_{\delta \tau, c}
  \quad , \quad
  \mathbf{r}=\mathbf{x}-\mathbf{x}^\prime,
}
we have
\eeq{
\mathcal C_n (\vec r) 
= {\tau_0^n} \sum_{m=1}^{n-1} \binom{n}{m} \xi^{(m,n-m)} (\vec r),
}
and thus
\eq{
\ln \Big( 1 + \la \delta F \delta F' \ra \Big)
  &= \sum_{n=2}^\infty \frac{(-1)^n}{n!} \mathcal C_n (\vec r) \\
  &= \sum_{n=2}^\infty {\tau_0^n} \Bigg[ \frac{1+(-1)^n}{2 (n/2)!}  \xi^{(n/2,n/2)} (\vec r)
+ 2 \frac{(-1)^n}{n!} \sum_{m=1}^{\floor{\frac{n-1}{2}}} \binom{n}{m} \xi^{(m,n-m)} (\vec r) \Bigg]. 
}
Note that the cumulants, $\xi^{(n,m)}$, contain the large--scale physics but the overall expression is complex which explains why the flux correlation function can look like linear theory on large scales but have complex, scale-dependent bias factors on smaller scales.

\subsection{Flux-quasar cross correlations}
\label{app:cross_correlations}

Let us briefly comment on the modeling of cross-correlations of the flux field with another tracer of large-scale structure, e.g.\ quasars or damped Ly$\alpha$ systems.  As we did for the flux auto-spectrum we shall neglect important observational issues such as continuum fitting, high column density systems, the proximity effect, etc.\ and focus on the intrinsic signal.  Thus we seek to model e.g.
\begin{equation}
    \left\langle (1+\delta_Q) e^{-\tau} \right\rangle
    = F_0 + \left\langle \delta_Q e^{-\tau} \right\rangle
    = F_0\left(1+\xi_{\times}\right)
\end{equation}
In perturbation theory we would write $\delta_Q$ as an expansion in powers of the matter overdensity, $\delta$.  We can proceed by considering the generating function $Z_{\times}(J,K) = \ln\left\langle \exp\left[J\tau + K\delta\right]\right\rangle$
where $\tau=\tau(\mathbf{x})$ and $\delta=\delta(\mathbf{x}^\prime)$.
As usual we define the cumulants\footnote{For this section we find it easier to use cumulants of $\tau$ rather than $\delta\tau$ as it leads to slightly more compact expressions.  The two differ only by powers of $\tau_0$ and a constant for the first cumulant.} as
\begin{equation}
    \kappa_{m,n} = \left. \frac{\partial^{m+n}}{\partial J^m \partial K^n}\ Z_{\times} \right|_{J=K=0}
\end{equation}
so that $\kappa_{m,n}$ is the coefficient of $J^m K^n/m!/n!$ in the expansion of $Z_{\times}$ and
\begin{equation}
    \left\langle e^{K\delta + J\tau} \right\rangle_{\delta,\tau} = \exp\left[ \sum_{m,n} \frac{J^m K^n}{m!n!} \kappa_{m,n}  \right] \quad .
\end{equation}
Taking $J=-1$ and considering $K$-derivatives of the above (then setting $K=0$) it is straightforward to show that
\begin{equation}
  \left\langle \delta^n e^{-\tau} \right\rangle
    = F_0  \sum_{m} \frac{(-1)^m}{m!}\ \kappa_{m,n}
  \quad , \quad
  F_0 = \exp\left[ \sum_m\frac{(-1)^m}{m!}\ \kappa_{m,0} \right]
  \quad .
\end{equation}
If we use this formula with the normal bias expansions for $\delta_Q$ and $\delta\tau$ the form of $\xi_\times$ quickly follows.


\section{Eulerian perturbation theory calculation}
\label{app:ept_calc}

In this appendix we give details of the redshift-space $\tau$ cumulants at one-loop order in (standard) Eulerian perturbation theory.  We include this calculation because the formalism may be familiar to many readers, and to make connection with similar calculations within this framework.

At linear order the second cumulant gives the usual super-cluster infall effect which in Fourier space takes $P\to (b_1+f\mu^2)^2P$ \cite{Kai87}. Beyond linear theory the second cumulant remains equivalent to the redshift-space correlation function of a biased tracer, so we will not repeat the expression here but refer readers to e.g.\ ref.~\cite{Chen20} and the previous appendix for the expression with the same biasing convention.  We note that some of these loop contributions are proportional to the linear $P(k)$ on large scales, modifying the amplitude of this term by factors depending upon small-scale physics.
Also, at one-loop order any correlator involving four powers of $\tau$ must be disconnected, so the fourth cumulant vanishes. This leaves us with the third cumulant, $\avg{\delta \tau_1 \delta \tau_2^2}$, which we now compute.  Our focus will be on the terms most sensitive to small-scale physics, the so-called UV-sensitive terms.

There are two contributions to the third cumulant quadratic in the linear power spectrum:
\begin{equation}
    \avg{\delta \tau_1 \delta \tau_2^2} = 2\avg{\delta \tau^{(1)}_1 \delta \tau^{(1)}_2 \delta\tau^{(2)}_2} + \avg{\delta \tau^{(2)}_1 \delta \tau^{(1)}_2 \delta\tau^{(1)}_2}.
\end{equation}
The first is
\begin{equation}
    \mathrm{FT} \left\langle\delta \tau_s^{(1)}(\mathbf{r}_1)\ \delta \tau_s^{(1)}\delta \tau_s^{(2)}(\mathbf{r}_2) \right\rangle =
    2 Z_1(\bk) P(k)\int\frac{d^3\bp}{(2\pi)^3}\ Z_1(\bp) Z_2(\mathbf{k},\bp)P(p).
\label{eqn:xi112}
\end{equation}
The first and second order redshift space kernels are given by \cite{Ber02}
\begin{equation}
    Z_1(\bk_1) = b_1 + f\mu_1^2
\end{equation}
and
\begin{align}
    Z_2(\bk_1, \bk_2) = b_1F_2(\bk_1,\bk_2) &+ f\mu_k^2 G_2(\bk_1,\bk_2) + \frac{b_2}{2} + b_s \left( \frac{(\bk_1\cdot\bk_2)^2}{k_1^2 k_2^2} - \third \right) \nonumber \\
    &+ \frac{fk\mu_k}{2} \left[ \frac{\mu_1}{k_1}(b_1+f\mu_2^2) + \frac{\mu_2}{k_2}(b_1+f\mu_1^2) \right]
\end{align}
where $\mu_k = \hk \cdot \hn$, $\bk = \bk_1 + \bk_2$ and $F_2$ and $G_2$ are the standard density and velocity kernels in Eulerian perturbation theory \cite{Ber02}:
\begin{equation}
    F_2(\bk_1,\bk_2) = \frac{5}{7} + \frac{1}{2}\frac{\bk_1\cdot \bk_2}{k_1k_2}\left(\frac{k_1}{k_2}+\frac{k_2}{k_1}\right)
    + \frac{2}{7}\frac{(\bk_1\cdot \bk_2)^2}{k_1^2k_2^2}
\end{equation}
\begin{equation}
    G_2(\bk_1,\bk_2) = \frac{3}{7} + \frac{1}{2}\frac{\bk_1\cdot \bk_2}{k_1k_2}\left(\frac{k_1}{k_2}+\frac{k_2}{k_1}\right)
    + \frac{4}{7}\frac{(\bk_1\cdot \bk_2)^2}{k_1^2k_2^2}
\end{equation}
To isolate the pieces sensitive to small-scale physics, let us take the $\bk \rightarrow 0$ limit of the integrand in Equation~\ref{eqn:xi112} (or more correctly consider $k\ll p$).  Dropping the terms that are odd in $\hp$ we get
\begin{equation}
    Z_2(\bk,\bp) \rightarrow \Bigg[ b_1\left(\frac{5}{7} + \frac{2}{7}\ [\hp \cdot \hk]^2 \right) + \left(\frac{3}{7} + \frac{4}{7}[\hp \cdot \hk]^2 + \frac{b_1}{2} + \frac{f\mu_k^2}{2}\right) f \mu_p^2 + \frac{b_2}{2} + b_s \left( [\hp \cdot \hk]^2 - \third \right) \Bigg] \nonumber.
\end{equation}
Using the angular averages
\begin{equation}
    \int\frac{d\Omega_p}{4\pi} (\hk \cdot \hp)^2 = \frac{1}{3}
    , \quad
    \int\frac{d\Omega_p}{4\pi} (\hk \cdot \hp)^2 (\hn \cdot \hp)^2 = \frac{1+2\mu_k^2}{15}
    , \quad
    \int\frac{d\Omega_p}{4\pi} (\hk \cdot \hp)^2 (\hn \cdot \hp)^4 = \frac{1+4\mu_k^2}{35}
    \label{eqn:ang_ints}
\end{equation}
Equation (\ref{eqn:xi112}) then becomes
\begin{align}
    2\sigma^2  (b_1 + f\mu_k^2) &\Bigg( \left[ \frac{5}{7}\left(b_1 + \frac{f}{3}\right) + \frac{2}{7}\left( \frac{b_1}{3} + \frac{1+2\mu_k^2}{15}f\right) + \frac{b_2}{2} \right] b_1 \nonumber \\
    &+ \left[ \frac{3}{7}\left(\frac{b_1}{3} + \frac{f}{5}\right) + \frac{4}{7} \left(\frac{1+2\mu_k^2}{15} b_1 + \frac{1+4\mu_k^2}{35}f\right) + \frac{b_2}{6} + \left(\frac{1+2\mu_k^2}{15} - \frac{1}{9} \right) b_s \right]f \nonumber \\
    &+ \frac{f}{2} (b_1 + f\mu_k^2) \left(\frac{b_1}{3} + \frac{f}{5}\right)\Bigg)\ P(k)
\end{align}
where
\begin{equation}
    \sigma^2 = \int\frac{p^2\,dp}{2\pi^2}  P(p)
    \quad .
\end{equation}
As $k\to 0$ this is of the form
\begin{equation}
    2\ (A_0 + A_2 \mu^2) (b_1 + f\mu^2)\ \sigma^2\,P(k)
\end{equation}
with
\begin{align}
    A_0 &= \frac{17b_1^2}{21} + \frac{46b_1f}{105} + \frac{b_1^2f}{6} 
    + \frac{5f^2}{49} + \frac{b_1f^2}{10} + \frac{b_1 b_2}{2} + \frac{f b_2}{6} - \frac{2 f b_s}{45} \, \\
    A_2 &= \frac{4b_1f}{35} + \frac{16f^2}{245} + \frac{b_1f^2}{6} + \frac{f^3}{10} + \frac{2f b_s}{15}\ .
\end{align}
Note that this has similar angular structure to the linear theory result, but the coefficients of the powers of $\mu$ are different and depend upon large momenta in the integral over $P$, i.e.\ on small-scale physics.

Similarly, the second contribution is given by
\begin{align}
    \mathrm{FT}\avg{\tau^{(2)} | \tau^{(1)} \tau^{(1)}} &= 2 \int \frac{d^3 p}{(2\pi)^3}\ Z_2(p, k-p) Z_1(-p) Z_1(p-k) P(p) P(k-p) \nonumber \\
    &\rightarrow {\rm ``const_0"} +
{\rm ``const^{(0)}_2"} \frac{k^2}{k_\star^2}
+ {\rm ``const^{(2)}_2"} \mu^2 \frac{k^2}{k_\star^2}
+ \ldots \quad (k \rightarrow 0)
\end{align}
Taking the same limit in this equation simply produces a constant at leading order in $\bk$, which leads to a $\delta$-function in configuration space.

\section{General structure in the large-scale limit}
\label{app:general_structure_k_to_0}

In this appendix we fill in the technical details of the argument in Section~\ref{sec:more_loops} and consider in general terms the effects of loop contributions beyond those enumerated in the previous appendix. We shall focus on those that renormalize the linear theory predictions while commenting briefly at the end on contributions beyond it.

\subsection{Renormalizing linear theory}

As in the main text let us write the order $n$ contribution to the $\delta \tau^m$ as
\begin{equation}
    \int_{p_1 ... p_n} K^{(m)}_n(p_1, ..., p_n)\ \delta_0(p_1) ... \delta_0(p_n)\ (2\pi)^3 \delta^D(k - \sum p_{i}).
\end{equation}
The contributions of interest come from those linking only one power of the linear field between two points (Fig.~\ref{fig:beyond_one_loop})
\begin{align*}
    \int_{k,p_i} \int_{k',q_i} &K^{(a)}_n(k, p_1, ..., p_{n-1}) K^{(b)}_m(k', q_1, ..., q_{m-1}) \\
    &\avg{ \delta_0(k) \delta_0(p_1) ... \delta_0(p_{n-1})\ (2\pi)^3 \delta^D(\sum p_i)\  \delta_0(k') \delta_0(q_1) ... \delta_0(q_{n-1})\ (2\pi)^3 \delta^D(\sum q_i) }.
\end{align*}
Contracting the $k, k'$ modes yields the contribution
\begin{equation}
    (2\pi)^3 \delta^D(k+k')\ \kappa^a_n(\bk)\ \kappa^b_m(\bk)\ P(k)
\end{equation}
where we have defined
\begin{equation}
    \kappa^{(a)}_n =  \int_{p_i}\ K^{(a)}_n(k, p_1, ..., p_{n-1}) \avg{\delta_0(p_1) ... \delta_0(p_{n-1})\ (2\pi)^3 \delta^D(\sum p_i)}
\end{equation}
which can depend on $\mu$ at leading order in $k$ as evident in our explicit calculation for $a, b = 1, 2$. That these contributions can all be so factored implies that the total contribution to the correlation function logarithm can be summarized by a single renormalized linear bias factor
\begin{equation}
    {\rm FT} \{ \ln(1 + \xi_{FF})\} \supset \sum_{a,b} \kappa^{(a)} \kappa^{(b)} P(k) = \Big(\sum_a \kappa^{(a)} \Big)^2\ P(k) \rightarrow b^2(\mu) P(k)
    \quad\mathrm{as\ } k \rightarrow 0
\end{equation}
where we have defined $\kappa^{(a)} = \sum_n \kappa^{(a)}_n$.

Let us now consider the structure of $b(\mu)$ by inspecting the large scale ($k \rightarrow 0$) limit. To do so we expand the kernels $K^{(a)}_n$ in $k$ to consider the $\mathcal{O}(k^0)$ contributions. Generically this will take the form
\begin{equation}
\lim_{k \rightarrow 0} K^{(a)}_n(k, p_1, ..., p_{n-1}) = a^{(0)} + a^{(2)}_{ij} \hk_i \hk_j  + a^{(4)}_{ijkl} \hk_i \hk_j \hk_k \hk_l + ...
\label{eqn:ang_struct}
\end{equation}
The coefficients $a^{(n)}$ come from integrating over the $p_i$ modes and, by the symmetries of the problem, must be factorable into products of the Kronecker $\delta$ and the line-of-sight direction $\hn$. They are defined to be order $\mathcal{O}(k^0)$ so are the relevant contributions in the large scale limit. Indeed we can see that the integrals in Equation~\ref{eqn:ang_ints} fall into this form and truncate at $a^{(2)}_{ij}$. The $\mu$ dependence of $b(\mu)$ therefore depends on at what order in $\hk$ this expansion truncates. As a first simplifying step we note that $K^{(a)}_n$ are products of redshift-space densities in configuration space, i.e. convolutions of $n^{\rm th}$ order redshift-space kernels in Fourier space, that is 
\begin{equation}
    K^{(a)}_n(k, p_1, ..., p_n) = Z_{n_1}(k, p_1, ..., p_{n_1-1})\ Z_{n_2}(p_{n_1}, ... )\ Z_{n_3}(p_{n_1+n_2},...)\ ...
\end{equation}
with conservation of momentum enforced by a delta function when integrating over $k, p_i$. Since $k$ only appears in one of the $Z_{n_i}$ it is sufficient to look at these independently.

In fact, the redshift-space kernels can themselves be decomposed into constituent density, velocity and bias kernels. Specifically we can write \cite{Ber02}
\begin{equation}
    \delta_s(\bk) = \sum_{n=0}^\infty \int_{\bk_i} \left[ \delta(\bk_1) + f \mu_1^2 \theta(\bk_1) \right] \frac{(f\mu k)^{n-1}}{(n-1)!} \frac{\mu_2}{k_2} \theta(\bk_2) ... \frac{\mu_n}{k_n} \theta(\bk_n) (2\pi)^3 \delta^D(\bk - \sum_i \bk_i).
\end{equation}
From this structure we can deduce that for each power of $\delta$  or $\theta$ either (a) they are linear modes with momentum $\bk_i$, in which case their $\hk_i$ dependence can either come from a $\mu_i^2$ or $\mu_i/k$ or (b) they are nonlinear modes with total momentum $\bk_i$ dominated by short modes with one long mode $\bp$ hidden inside, in which case the angular dependence on the long mode depends solely on the nonlinear kernel of $\delta$ or $\theta$ since $\mu_i$ doesn't depend on $\bp$ to lower order.

Indeed, the nonlinear bias and matter kernels also satisfy this structure. Let us begin with the latter as an instructive example. The nonlinear matter density and velocity kernels $F_n$ and $G_n$ obey recursion relations that take the schematic form \cite{Ber02}
\begin{equation*}
    F_n(\bq_1,...,\bq_n) \sim G_m(\bq_1,...,\bq_m) \Big( \alpha(\bk_1,\bk_2) F_{n-m}(\bq_{m+1},...,\bq_n) + \beta(\bk_1,\bk_2) G_{n-m}(\bq_{m+1},...,\bq_n) \Big)
\end{equation*}
where we have omitted the implied sum over $m$ and numerical factors, and defined \cite{Ber02}
\begin{equation}
    \alpha(\bk_1,\bk_2) = \frac{\bk_{12}\cdot\bk_1}{k_1^2}
    \quad , \quad
    \beta(\bk_1,\bk_2) = \frac{ k_{12}^2 (\bk_1 \cdot \bk_2)}{2 k_1^2 k_2^2}
    \quad ,
\end{equation}
with $\bk_1= \bq_1 + ... + \bq_m,$ $\bk_2 = \bq_{m+1} + ... + \bq_n$ and $\bk_{12} = \bk_1 + \bk_2$.
The velocity kernels, $G_n$, have the same form. We are interested in the limit where one of the momenta ($\bp$) is much smaller than the others, such that it can be ignored when summed with other momenta, $\bq_i$. In this limit we see that the leading contribution of $\bp$ to the denominators of $\alpha$ and $\beta$ are at most order $\mathcal{O}(p^2)$ when $m = 1, n-1$, and $\mathcal{O}(p^0)$ otherwise. However in both of these cases they multiply $F_1, G_1 = 1$. Thus in general the terms contributing to kernels at order $n$ have denominators $1/p^2$ coming from $\alpha,\beta$ or from the $F$'s and $G$'s themselves, which do not produce additional orders when multiplied because they carry different momenta. This implies that each term has at most $p^2$ in the denominator. This then implies that expansion as in Equation~\ref{eqn:ang_struct} must terminate at $p_i p_j/p^2 = \hat{p}_i \hat{p}_j$. Recently, \cite{Fujita+:2020} showed that all real-space bias operators can be expressed at each order by combining lower-order operators $X, Y$ though
\begin{equation}
    XY, \; \Big(\partial_i X\Big) \Big(\frac{\partial_i}{\partial^2} Y\Big), \Big(\frac{\partial_i \partial_j}{\partial^2} X\Big) \Big(\frac{\partial_i \partial_j}{\partial^2} Y\Big)
\end{equation}
e.g.\ the quadratic shear bias can be formed by applying the third combination to the linear operator $\delta$. From a similar argument about joining operators that we used in the redshift-space formula above we can conclude that no real-space bias operator can produce a dependence on long modes more complicated than second-order in $\hk$ using just the structure of dynamical nonlinearities in structure formation.

\subsection{Two and more soft modes}

We can also extend the arguments of the previous subsection to the response of the flux field to more than one long-wavelength mode. To do so it is necessary to consider the $n^{\rm th}$ order kernel above with more than one long-wavelength mode, i.e.
\begin{equation}
    \kappa^{(a)}_n(\bk_1,\bk_2) =  \int_{p_i}\ K^{(a)}_n(k_1,k_2,, p_1, ..., p_{n-2}) \avg{\delta_0(p_1) ... \delta_0(p_{n-m})\ (2\pi)^3 \delta^D(\sum p_i)}.
\end{equation}
We can consider two cases. In one the two soft modes can eventually be reduced to two different redshift-space kernels, i.e.\ $Z_{n_1}(\bk_1,p_1, ...) Z_{n_2}(\bk_2, p_2,...)$ in which case previous arguments imply factorizable forms like 
\begin{equation}
    \delta(\bp)\ \frac{q_z q_z}{q^2} \delta(\bq), \; \frac{p_z p_z}{p^2} \delta(\bp)\ \delta(\bq), \; \frac{p_z p_z}{p^2} \delta(\bp)\  \frac{q_z q_z}{q^2} \delta(\bq).
\end{equation}
However, these two linear modes can also be coupled by nonlinearities. For this argument it will be useful to write down the general form of anisotropic couplings induced by redshift-space distortions; from Section 3 of ref.~\cite{Ivanov2018} we can see that these generally take the form of
\begin{align}
    \alpha^s(\bk_1,\bk_2) &= \frac{k_{1,z} (k_{1,z} + k_{2,z})}{k_1^2} \sim X\ \hn_i \hn_j \frac{\partial_i \partial_j}{\partial^2} Y, \; (\hn_i \partial_i X)\ \hn_j \frac{\partial_j}{\partial^2}Y \\
    \beta^s(\bk_1,\bk_2) &= \frac{(k_1 + k_2)^2 k_{1,z}k_{2,z}}{2 k_1^2 k_2^2} \sim (\hn_i \partial_i X)\ \hn_j \frac{\partial_j}{\partial^2}Y,\; \Big(\frac{\partial_i \partial_{\hn}}{\partial^2} X \Big) \Big(\frac{\partial_i \partial_{\hn}}{\partial^2} Y \Big).
\end{align}
There are essentially two cases of this---they can be couple by real-space nonlinearities, with anisotropy introduced ``in front'' e.g.\ $\mu_{p+q}^2 G_n(p,q,...)$ or directly coupled in an anisotropic way i.e.\ $\hp_i \hp_z \delta(\bp)\ \hq_i \hq_z\delta(\bq)$. At quadratic order this recovers the terms in ref.~\cite{Givans20}, though we have neglected displacement effects due to operators in the form of $\partial_i/\partial^2$, whose coefficients can be fixed by equivalence-principle considerations \cite{Fujita+:2020}. We intend to return to this topic in a future work.

\bibliographystyle{JHEP}
\bibliography{main}
\end{document}